# The Cosmic Ray Hodoscopes for Testing Thin Gap Chambers at the Technion and Tel Aviv University

E. Etzion[1,2], H. Abramowicz[2], N. Amram[2], Y. Benhammou[2], M. Ben-Moshe[2], G. Bella[2],
J. Ginzburg[2], Y. Gernitzky[3], A. Harel[3], H. Landsman[3], N. Panikashvili[3], Y. Rozen[3],
S. Tarem[3], E. Warszawski[3], J. Wasilewski[4], L. Levinson[5],

*Abstract*--**Thin gap chambers (TGCs) are built for the muon trigger chambers in the endcap region of the LHC experiment ATLAS. More than 2500 ATLAS TGCs are being produced at the Weizmann institute in Israel, and in Shandong University in China. Detailed testing of these chambers is performed at the Technion and at the Tel-Aviv University. Two cosmic ray hodoscopes for testing the operation of these detectors were built in Israel. In these hodoscopes the response of the chambers to energetic cosmic ray muons is recorded and analyzed. The hodoscopes measure the exact time and space location of the cosmic ray hit and read out the chambers which are being tested to verify that they produce a corresponding signal within the required time interval. The cosmic ray hodoscopes built at the Technion and at the Tel Aviv University for the test of ATLAS TGCs are described. The mechanical structure, readout electronics, data acquisition and operating scheme are presented. Typical TGC test results are presented and discussed.**

Manuscript received July 1, 2003. This work is supported by the Israel Science Foundation and the German Israeli Foundation.

1. CERN, European Organization for Nuclear Resaerch, CH-1211 Geneve 23 Switzerland (telephone: +41227671153, e-mail: erez.etzion@cern.ch).

2. School of Physics and Astronomy, Raymond and Beverly Sackler Faculty of Excat Sciences, Tel Aviv University, Tel Aviv 69978, Israel (telephone: +97236407722).

3. Physics Department, Technion, Haifa 32000, Israel (telephone: +97248963553).

4. Tel Aviv University, currently at the Technical University of Lodj, Poland

5. Particles Physics Department, Weizmann Institute of Science, Rehovot 76100, Israel (telephone: +97289342084).

## I. INTRODUCTION

THE Large Hadron Collider (LHC) is a new proton-proton collider under construction at the European Center for Nuclear Research (CERN). Two 7 TeV proton beams will collide at the highest energy ever reached in a particle accelerator with the design luminosity of $10^{34}$ cm$^{-2}$s$^{-1}$. The ATLAS detector at LHC expects to receive beam events at a rate of up to $10^9$ Hz. The trigger system is designed to reduce the event rate to about 100 Hz. Amongst the primary goals of the experiment are understanding of fundamental symmetry breaking which might manifest itself in the observation of the Higgs boson and possibly new phenomena or new objects. Therefore while requiring an overall rejection of about seven orders of magnitude, an excellent triggering efficiency must be retained for the detection of rare new physics processes. The trigger equipment [1] accepts data at the full LHC bunch-crossing rate of 40 MHz (every 25 ns) and must identify the bunch crossing containing the interaction of interest and introduce negligible dead time.

ATLAS is a cylindrical shape four $\pi$ detector, 44 m long and 22 m high. The TGCs [2] are dedicated to form a muon trigger in ATLAS two end-cap regions. The Seven TGC layers consist of one triplet (three wire planes) and two TGC doublets (each consists of two wire planes). The chambers are mounted on big wheels in the end-cap region about 14 m away from the interaction point [3]. A TGC consists of a plane of closely spaced wires maintained at positive high voltage (HV), sandwiched between resistive grounded cathode planes. A TGC is characterized by an anode wire to cathode plane gap distance which is smaller than the wire-to-wire spacing. ATLAS TGCs are produced with 1.4 mm wire to cathode plane distance and 1.8 mm for the wire to wire spacing. The sizes of TGC units depend on their position and vary between 1.31 m$^2$ to 2.27 m$^2$.

The TGCs are well suited for triggering the high transverse momentum muons at the LHC as they operate in a saturated proportional mode, giving fast signals with a typical rise time below 5 ns. The saturated proportional mode is gained under a strong electric field obtained with 50 microns gold plated tungsten wires and a very narrow gas gap. The shower amplifications and quenching are optimized with a mixture of 45% n-pentane and 55% $CO_2$. This operation mode leads to strong signals with reduced Landau tails and a high signal to noise ratio. To form a trigger signal, between 4 to 20 anode wires are grouped together and fed to a common readout channel. The wire and strip signals emerging from the TGCs are fed into a two-stage amplifier in an Amplifier Shape Discriminator (ASD) circuit [4]. Four such circuits are built into a single ASD chip and four ASD chips are incorporated into an ASD Board.

Three production lines build the TGCs for ATLAS. The construction of the 3,600 chambers is shared between Weizmann Institute of Science (Israel) building 2,160 chambers, KEK (Japan) building 1,056 chambers and Shandong University (China) building 384 chambers. The chambers built in Japan are tested in a testing facility in Kobe University [5], the chambers built in Israel and China are tested in the two cosmic ray (CR) hodoscopes constructed for that purpose in the Technion and Tel Aviv University.

## II. COSMIC RAY MUONS

Most of the CR muons are produced high in the atmosphere (typically at 15 km above earth) by the decay of charged mesons. They reach the ground with a mean energy of approximately 4 GeV [6]. The integral intensity of vertical muons above 1 GeV/c at sea level is approximately 70 m$^{-2}$ s$^{-1}$ sr$^{-1}$, which can be written in the form I $\approx$ 1 cm$^{-2}$ min$^{-1}$ for horizontal detectors. The overall zenith angle distribution of muons at the ground is proportional to $\cos^2\theta$.

## III. THE TESTBENCH GENERAL STRUCTURE

The hodoscopes use cosmic muons to measure the time response and detection efficiency of the TGCs. The two systems are similar but minute differences exist. Hence, the following is an accurate description of the Tel Aviv system [7], and reliably portray the Technion system [8] as well. Up to eight TGC units (doublets or triplets) which are tested in parallel are interleaved between two precision chambers (PRC) and two layers of triggering scintillators. A Schematic structure of the testbench is shown in Fig. 1. A picture of one of the hodoscopes is shown in Fig 2.

The information that a muon crossed the PRC, and therefore the tested stack, is provided by two scintillator planes, one above the upper PRC and one below the lower PRC. Each scintillator plane consists of four scintillator slabs, disconnected optically, 60 cm wide, 140 cm long, and 1.2 cm thick. The scintillator material has a high index of refraction (n=1.58), and it is polished accurately to have total reflection, not allowing light produced inside to escape. The emittance spectrum of the scintillator material is in the visible range.

The two ends of each scintillator slab are glued to triangular light-guides that guide the collected light to respective Photo multipliers (PMT).

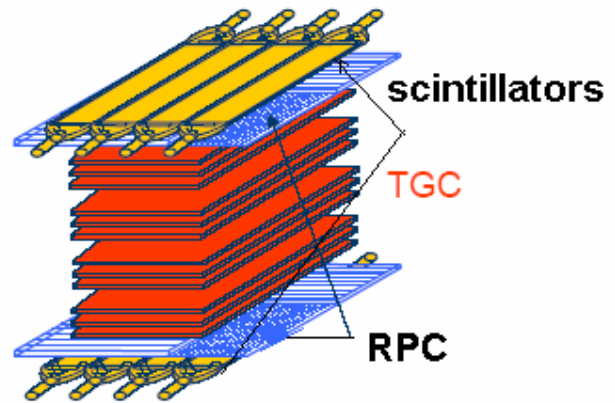

Fig. 1 A schematic structure of the CR hodoscopes in the Technion and Tel Aviv University. The tested TGC units are interleaved within triggering layers (the scintillators) and position measurement units (the precision chambers).

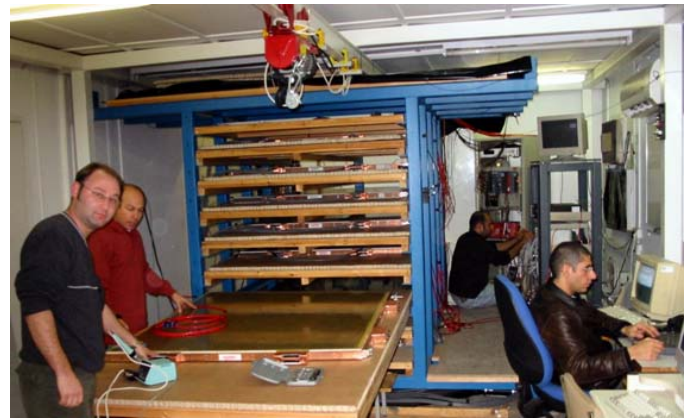

Fig. 2 A photograph of the testbench at Tel Aviv.

We use Hamamatsu R329-02 PMTs, which are active in the range of 300-650 nm and have a rise time of 2.6 ns. A coincidence between a hit in the upper and the lower scintillator planes is used to trigger the data taking.

For calculating the hit position we use the readout of dedicated Precision Chambers (PRC). The PRCs are special rectangular shape TGCs, which consists of two perpendicular layers of strips and one layer of HV wires between them. There are 361 strips along the X-axis of the rectangle, and 458 strips along its Y-axis. Each strip has a width of 3.6 mm. The signals from the two layers are latched and read sequentially providing the X and Y coordinates of a hit position. Along the wires the hit position is derived by using the charge distribution between adjacent strips, this gives us a resolution of a few hundred microns. In the transverse direction the hit is found at the closest strip, resulting in a precision of 1.8 mm.

The precision chambers and the TGC detectors need a constant gas flow. The gas system produces a mixture of 45% n-pentane and 55% of $CO_2$, which prevents discharges in the chambers. This right mixture is achieved by bubbling $CO_2$ gas through liquid n-pentane at temperature of a $17^0$ C. Two gas lines supply gas from the gas mixing system. One line goes to the chambers in the testbench and the other to the batch of chambers in the pre-test stage. A single gas supply line carries about 20 l/h of the gas mixture.

The TGCs and the PRCs operate at 2.9-3.0 kV. This is supplied by a CAEN 40 channels SY127 HV unit. The SY127 supply also the negative HV required by the scintillators PMTs. The SY127 is locally monitored and controlled via on-board alpha-numeric keypad and display. It is also remotely controlled via RS-232-C port and the VME Controller mod. V288.

The electronic readout system is mounted on a two racks equipped with standard NIM crate and a VME crate.

## IV. READOUT SCHEME

The following electronic devices are utilized for reading the TGC and PRC chambers:

- TGC readout – The TGCs are read via the ASD IC and 16 channel ASD boards [4]. These ICs, using SONY master slice bipolar technology, are the ones that were designed and build for the TGC readout in ATLAS. The ASDs are attached to the edge of a TGC chamber and enclosed inside the TGC electrical shielding.
- PRC readout – Gassiplex [9] chip is a low noise analog readout processor which was developed at CERN to read and amplify the signal from gaseous detectors. Each of the PRC strips is connected to the Gassiplex via a high currents protection card.

A VME bus is used as a common interface between the "online computer" which controls the measurement and stores the data, and the lab measuring equipment. The following cards are connected through the VME bus:

- A CPU module - a 1GHz Pentium III runs a Linux operating system, with 256 MB SDRAM. It includes all the standard PC I/O (keyboard, mouse, SVGA, IDE, FDC, COM1/2, LPT1), a Flat Panel, 10/100BaseTX, Fast/Wide SCSI-2, optional transition module CD-ROM/floppy and IDE hard drive; and in addition a VME64 support.
- TDC – CAEN V767 is a multi-hit time to digital converter (TDC), a one-unit wide VME module, houses four chips, each with 32 time to digital conversion channels (128 channels in total).
- C-RAM – CAEN module V550 is an analog to digital converter (ADC). C-RAM is a one-unit wide VME module houses two independent analog to digital conversion blocks.
- SEQUENCER – CAEN module V551B is a one-unit wide VME module that handles the Data Acquisition from the PRC. The SEQUENCER has been developed to control the signals from/to the C-RAM boards.
- HV – CAEN module V288 is a High Speed VME Controller Interface. This unit controls the HV SY127 distribution of power to the scintillators and the TGC and PRC chambers.

The following NIM cards are used for the signal processing and manipulations and logics decisions:

- CAEN NIM - TTL adapter module N89 houses two sections of four NIM to TTL converters and two sections of four TTL to NIM converters.
- CAEN NIM - ECL adapter converts the NIM standard signals to ECL standard signals.
- ASD readout board and ASD-TDC units that were designed by the Weizmann institute electronic and data acquisition unit. These cards read the signals from the on-chamber ASD boards, and send the signals to the TDC card in ECL format. Each of the two boards in these modules receives 32 signals in LVDS format from two ASDs and output two differential ECL signals. Each output signal represents the OR of 16 odd or 16 even wire or strip signals.
- LeCroy constant fraction discriminator generates a precise digital logic pulse, from an analog one, if the latter exceeds a given threshold.
- LeCroy OR/AND logical units generates an AND or an OR signal between two discriminator generated signals.

The signals from the scintillators are used to trigger the system and set a time reference to the measurements. The eight PMT signals of each scintillator plane are connected via the discriminators to an OR unit. The output of the two OR units are then passed to an AND logical unit, which its output serves as a trigger signal for the readout system.

The signals from the OR units, delayed by 100 ns go through the NIM-ECL converter to the TDC unit, via the ASD-TDC card.

The signals from the TGCs go through the ASD, the ASD readout board and the ASD-TDC converter to the TDC for the TGC time response measurement. Our study showed that more than 95% of the signals arrived in a time window shorter than 24 ns with an uncertainty of 1.5 ns.

The PRC signals are serially read through the Gassiplex to the ADC unit. It is controlled by the SEQUENCER using CLOCK HOLD and CLEAR signals. The CLOCK determines the time length for reading one channel. The HOLD signal freezes the information that is currently in the Gassiplex chip; the CLOCK pushes the signal to the output (C-RAM) and then a CLEAR signal resets all the Gassiplex chips. The readout process takes 2-4 microsecond per channel, but the channels are read in parallel into several ADCs.

The electronic connections between all the different units in the readout systems are schematically depicted in Fig. 3

## V. ANALYSIS METHOD

The Quality Assurance (QA) of the chambers produced in Israel begins at the production site [10].
All the chambers are flushed with $CO_2$ and run with 2.9 kV for one day. After that the gas is changed to the operation mixture of $CO_2$ and n-pentane (45%-55% respectively) and the chambers are kept two days with that mixture. Then it is operated with the nominal voltage (3 kV) and 3.2 kV. After checking that the entire channels are operative the chambers are tested with radioactive source searching for potential noise problems. Each detector is irradiated with 2500 Ci $CO_{60}$ source at a nominal operating voltage and at photon rate several times above the expected background (for further details [11]). Scanning with the radiation source is used to detect hot spots in the chambers.

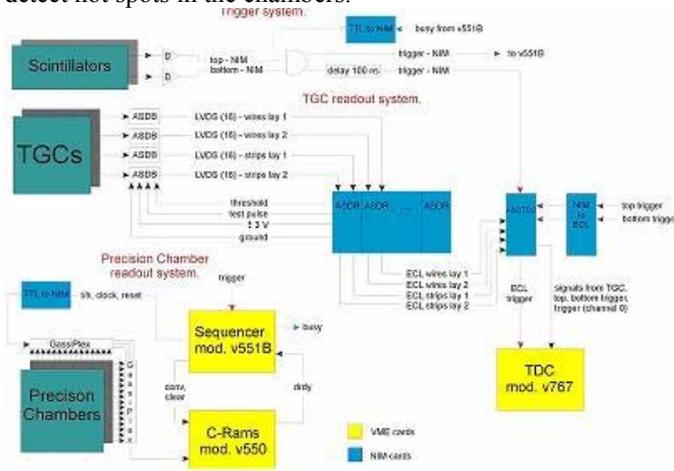

Fig. 3 The layout of the readout system

Chambers that pass these preliminary tests are delivered to Tel Aviv or the Technion for further detailed check.
The QA in the test laboratories is as follows:

### A. *Preliminary check*
Incoming TGC units are identified by their barcode number and are registered in the database (DB). MySQL DB serves an electronic logbook for the tests. Information is stored and results are presented via a Web graphic user interface. The DB is also directly connected to the offline software and feeds it with the conditions of the testing runs. After registration the chambers go through the following tests:
1. Mechanical integrity inspection - A visual inspection of the incoming TGC units is performed to check that no mechanical damage was done.
2. Gas integrity of chambers - Each TGC unit, that passed the mechanical inspection, is flushed with $CO_2$ at 40 ml/min, for one day. Then, it is flushed with the operating gas ($CO_2$-n-pentane mixture) for additional two days.
3. HV check - After 48 hours of flushing the gas mixture, the module is tested under HV. The HV is first ramped to the nominal 2.9 kV. If no HV trip is observed, the HV is further increased to 3 kV. At this voltage, the current of each counter should be below $15 \mu A$. No sparking should be encountered. The current and number of sparks (if any) are recorded in the DB.

### B. *Cosmic Ray efficiency test*
Units that pass the preliminary tests are moved to the CR efficiency testbench; simultaneously the gas HV studies of the next batch are performed. The scope of the efficiency test is to measure the time response and the efficiency of the TGC counters. An online Data Acquisition (DAQ) program, running on the VME based Pentium CPU, is controlling the test. Every DAQ run begins with readout of 2,000 events with software generated random trigger. This is used to set the reference noise level threshold for each of the ADC channels. When a trigger is detected the DAQ reads the data from the TDC, the ADC (C-RAM), and the HV controller, and stores it in a binary output file for further offline analysis. In parallel the program continuously plots a set of histograms which enable to online monitor the quality of the data and the equipment performances. The histograms are automatically stored on the DB and are online accessible via the WEB.
For safety reasons a slow control program running on a separate computer serves as a watch dog and sends SMS warnings messages when its sensors detect changes in parameters like room temp, water temp or unexpected problems in the HV system.
Events accumulation for a period of about four days, allows a full mapping of the efficiency of each detector in the stack, to a precision of 1%.

### C. *Validation procedure*
A good chamber must have at least 95% of its active area more than 95% efficient. The time response should be less than 25 ns.
A C++ offline analysis program reads the binary output of the DAQ code and processes it. The offline program first reads the threshold parameters of the ADC channels. Then it sequentially reads the information stored for each muon crossing. Combined with the geometric setup of the chambers in the stack, which is directly read from the DB, the program can calculate the expected position of the muon intersection with every chamber and check if within 25 ns relative to the trigger a hit was detected by the wires and/or the strips of each chamber. The results of these comparisons are two dimensions efficiency histograms of the tested chambers.
The hit position in the PRCs is calculated as a weighted average of the ADC pulse height in each coordinate. A single muon signal is typically spread over four to ten strips in each coordinate. The program requires one cluster of adjacent strips in each of the four coordinate (X and Y on top and bottom PRCs). A cluster is considered as a good muon hit only if 90% of its charge is concentrated in the center of the cluster +/- five close neighbors. $\bar{x}_W$ in (1) represents the position calculated in the PRC and its error, where $x_i$ and $p_i$ are the position and the pulse height of a specific channel $i$, $P_{sub}$ is the pulse height contained within a restricted range of ±five channels from the peak and $\Delta P_i$ depends on the fluctuation of the signal in a single strip.

$$\overline{x}_W = \frac{\sum_i x_i P_i}{P_{sub}} \quad \pm \quad \sqrt{\sum_i (x_i - \overline{x}_W)^2 \cdot (\Delta P_i / P_{sub})^2}$$

(1)

Events with more than one cluster in one of the coordinates are rejected. The muon trajectory is assumed to go in a straight line between the two calculated intersections with the PRCs. This assumption is used to interpolate the position of the intersection with the scintillators and with each of the tested chambers.

Comparison between the weighted average and the expected Gaussian distribution showed a difference of less than 2 mm. A study of the measurement error as defined in (1) resulted in an error below 1 mm.

Additional error could occur from multiple scattering of the muon traveling through the chambers. This effect was studied with a dedicated Monte Carlo program which simulated the interaction of minimum ionizing particle going through a set of TGCs. It showed that the deviation from the straight line assumption at the TGC layers is at the level of 2 mm. This all goes well with the 1x1 cm$^2$ granularity we choose for our measurement resolution.

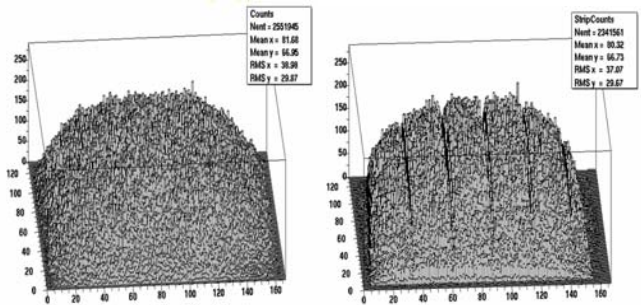

Fig. 4 Left: Expected distribution of TGC hits where the position is extrapolated from the hit position in the precision chambers. Right: Measured distribution of TGC hits. The clear difference seen between the two plots is the low efficiency lines on the right plot which are the result of the inactive area in the TGC support lines.

Fig. 4 presents the two dimensions TGC hit histograms, which are used to calculate a chamber efficiency plot. Plotted on the left side is the distribution of the number of hits expected in one of the TGCs as extrapolated from the hits in the two precision chambers. On the right hand side we have the same histogram but counting only the hits which had also a signal detected in the strips of that chamber. The granularity of these histograms is 1x1 cm$^2$. The efficiency map is the results of a bin by bin division of the above two histograms. An example of a strips efficiency map of one of the chambers is shown in Fig. 5.

One can clearly see in the figure the active area of the chamber, (the white area) as well as the five support lines and the 60 support buttons.

Separate plots are produced for the chamber's wires and strips; hence for a doublet it provides two wires and two strips efficiency plots. For a triplet there is an additional plot for the wires in the middle.

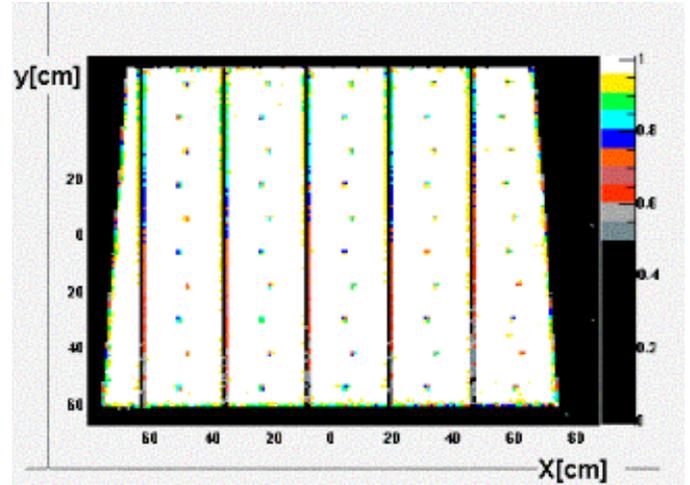

Fig. 5 A strips efficiency map of one of the chambers. The bin efficiency level is color coded with white standing for efficiency>95% and black for efficiency below 50%. The scale is given on the right side of the plot.

To quantify the quality of a chamber it was decided to use the following criteria:

- Inefficient area is defined as a contiguous area of more than 25 cm$^2$ that extends by more than five cm in each direction (x & y), with every point in that area having an efficiency of less than 95%.
- The decision if a chamber passes the test is based on the integrated size of surface of all the dead or inefficient areas in a detector.

To demonstrate the rules above see Fig. 6. The plot shows the strips efficiency distribution of the second chamber in the unit shown in Fig. 5. The main clear difference is that this chamber has a low efficiency region on one of its corners. The program has counted it as two small contiguous areas with a total inefficient region of 2.96%.

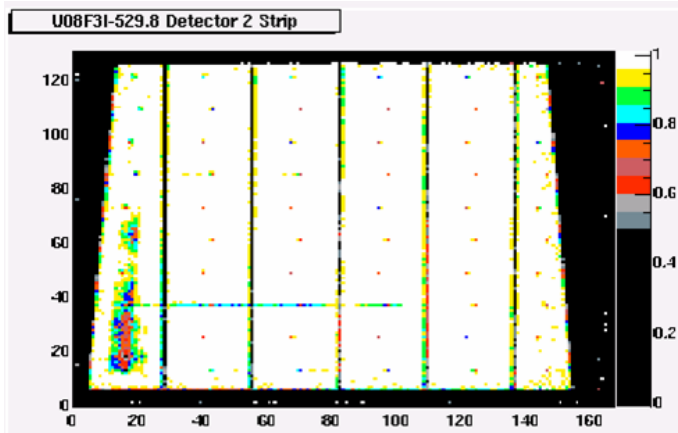

Fig. 6 An efficiency map of a chamber with low efficiency region on its left low corner.

Figure 7 presents the distribution of the total inefficient region in the so called T6 and T8 type chambers tested in Israel. It clearly demonstrates that the majority of the tested chambers passed the criterion of inefficient region below 5% of the active area.

The plot does not include chambers which were returned to Weizmann for well defined problems that can be fixed such as disconnected strips or wire groups.

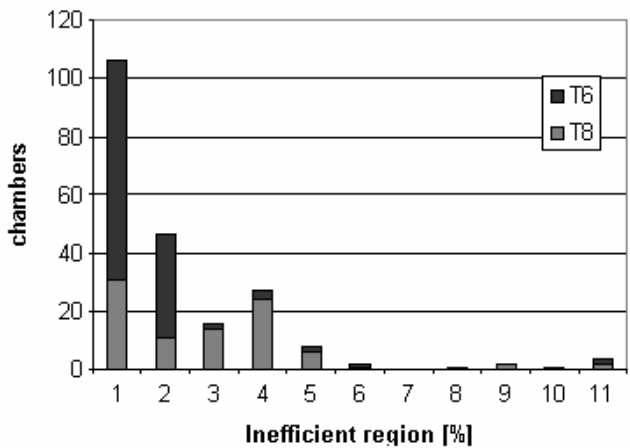

Fig. 7 A distribution of the total inefficient regions for the T6 and T8 chambers tested in Israel.

VI. SUMMARY

TGCs are tested in three different CR testing laboratories one in Japan and two in Israel, all using similar testing facility, and common testing schemes.

The CR test is a critical stage in the QA procedure of the chambers and successfully detects problematic chambers, providing efficiency map of each chamber and suggests operational parameters such as voltage and readout thresholds.

The laboratories at the Technion and Tel Aviv have tested already more than 300 T8 Units and 40 T6 type units (both doublets and triplets). The first two T9-type chambers which were produced in Shandong University were already successfully tested, and additional batch of 60 units will be shipped to Israel next month. About 20% of the Israeli chambers quotas are already at CERN after they completed all the test stages in Israel.

The results of the tests are stored in the DB which is accessible via standard WEB tools.

VII. ACKNOWLEDGMENT

We thank our colleagues from the ATLAS group of the Weizmann Institute of Sciences. We thank Z. Tarem for his contribution to the software design.

VIII. REFERENCES

[1] ATLAS Collaboration, "ATLAS High Level Trigger Technical Design Report", CERN/LHCC/90-014, 1998.
ATLAS HLT/DAQ/DCS Group, "ATLAS High Level Trigger Data Acquisition and Controls Technical design Report", http://atlas-proj-hltdaqdcs-tdr.web.cern.ch/atlas-proj-hltdaqdcs-tdr/tdr-v1-r4/PDF/TDR.pdf, 2003.

[2] S. Majewski et al., "A thin multiwire chamber operating in the high multiplication mode", Nucl. Inst. Meth. 217, 1983, pp. 265.
G. Mikenberg, "Thin Gap gas chambers for hadronic calorimetry", Nucl. Inst Meth. A265, 1988, pp. 223.
D. Lazic et al., "Measurement of drift velocity in several n-pentane based gas mixture and the influence of working gas composition on timing properties of Thin Gap Chambers", Nucl. Inst. Meth A410, 1998, pp. 159.
Mincer et al., "Chareg Production in Thin Gap Multiwire chambers", Nucl. Inst. Meth. A439, 2000, pp. 147-157.

[3] ATLAS Muon Spectrometer Technical Design Report, CERN/LHCC/97-22, 1997.

[4] O. Sasaki, "ATLAS Thin Gap Chamber Amplifier Shaper Discriminator ICs and ASD Boards", ASD PRR, 1999, http://online.kek.jp/~sosamu/ASD-PRR.pdf

[5] T. Sugimoto and the ATLAS Japan TGC group, "Cosmic Ray test system for the Thin Gap Chamber modules at Kobe", this proceedings.

[6] K. Hagiwara et al., *The Review of Particle Physics*, Phys. Rev. D66, 2002

[7] J. Wasilweski, "Data acquisition system for quality tests of the ATLAS muon endcap trigger chambers", MSc Thesis Technical University of Lodj, Tel Aviv, July 2002.
J. Ginzburg "Analysis and quality tests of the ATLAS muon endcap trigger chambers", MSc Thesis Tel Aviv University, January 2003.

[8] Y. Gernitzki et al., "The cosmic ray hodoscope for testing Thin Gap Chambers at the Technion", ATL-MUON-2003-005, CERN, July 2003.

[9] J.C Santiard et al., CERN-ECP/94-17, 1994.

[10] ATLAS TGC Collaboration, "Thin Gap Chambers Construction Manual", http://atlas.web.cern.ch/Atlas/project/TGC/www/TGC_cons_man.pdf

[11] V. Smakhtin and the ATLAS TGC Group, "General gamma-radiation test of TGC detectors", this proceedings.